\documentstyle[12pt]{article}
\def\Re{{\cal R \mskip-4mu \lower.1ex \hbox{\it e}\,}}
\def\Im{{\cal I \mskip-5mu \lower.1ex \hbox{\it m}\,}}
\def\ie{{\it i.e.}}
\def\eg{{\it e.g.}}

\def\sub#1{_{\lower.25ex\hbox{$\scriptstyle#1$}}}
\def\sul#1{_{\kern-.1em#1}}
\def\sll#1{_{\kern-.2em#1}}
\def\sbl#1{_{\kern-.1em\lower.25ex\hbox{$\scriptstyle#1$}}}
\def\ssb#1{_{\lower.25ex\hbox{$\scriptscriptstyle#1$}}}
\def\sbb#1{_{\lower.4ex\hbox{$\scriptstyle#1$}}}

\def\to{\rightarrow}
\def\mh{\ifmmode m\sbl H \else $m\sbl H$\fi}
\def\mch{\ifmmode m_{H^\pm} \else $m_{H^\pm}$\fi}
\def\mt{\ifmmode m_t\else $m_t$\fi}
\def\mc{\ifmmode m_c\else $m_c$\fi}
\def\mz{\ifmmode M_Z\else $M_Z$\fi}
\def\mw{\ifmmode M_W\else $M_W$\fi}
\def\mws{\ifmmode M_W^2 \else $M_W^2$\fi}
\def\mhs{\ifmmode m_H^2 \else $m_H^2$\fi}
\def\mzs{\ifmmode M_Z^2 \else $M_Z^2$\fi}
\def\mts{\ifmmode m_t^2 \else $m_t^2$\fi}
\def\mcs{\ifmmode m_c^2 \else $m_c^2$\fi}
\def\mchs{\ifmmode m_{H^\pm}^2 \else $m_{H^\pm}^2$\fi}
\def\ztwo{\ifmmode Z_2\else $Z_2$\fi}
\def\zone{\ifmmode Z_1\else $Z_1$\fi}
\def\mtwo{\ifmmode M_2\else $M_2$\fi}
\def\mone{\ifmmode M_1\else $M_1$\fi}
\def\tb{\ifmmode \tan\beta \else $\tan\beta$\fi}
\def\xw{\ifmmode x\sub w\else $x\sub w$\fi}
\def\ch{\ifmmode H^\pm \else $H^\pm$\fi}
\def\lum{\ifmmode {\cal L}\else ${\cal L}$\fi}
\def\inpb{\ifmmode {\rm pb}^{-1}\else ${\rm pb}^{-1}$\fi}
\def\infb{\ifmmode {\rm fb}^{-1}\else ${\rm fb}^{-1}$\fi}
\def\epem{\ifmmode e^+e^-\else $e^+e^-$\fi}
\def\ppb{\ifmmode \bar pp\else $\bar pp$\fi}

\newskip\zatskip \zatskip=0pt plus0pt minus0pt
\def\matth{\mathsurround=0pt}

\def\atversim#1#2{\lower0.7ex\vbox{\baselineskip\zatskip\lineskip\zatskip
  \lineskiplimit 0pt\ialign{$\matth#1\hfil##\hfil$\crcr#2\crcr\sim\crcr}}}

\renewcommand{\thefootnote}{\fnsymbol{footnote}}

\begin{document} \begin{titlepage}
\setcounter{page}{1}
\thispagestyle{empty}
\rightline{\vbox{\halign{&#\hfil\cr
&ANL-HEP-PR-93-20\cr
&April 1993\cr}}}
\vspace{1in}
\begin{center}

{\Large\bf
A Model of Universality Violation Revisited}
\footnote{Research supported by the
U.S. Department of
Energy, Division of High Energy Physics, Contracts W-31-109-ENG-38.}
\medskip

\normalsize THOMAS G. RIZZO
\\ \smallskip
High Energy Physics Division\\Argonne National
Laboratory\\Argonne, IL 60439\\

\end{center}

\begin{abstract}

The possibility that the interactions of the third generation of quarks and
leptons may violate universality by a small amount remains an open
experimental question. The model of Li and Ma, which naturally accommodates
such violations, is found to be highly constrained by newly obtained, high
precision electroweak and $\tau$-lepton data once full Standard Model
radiative corrections are incorporated into the analysis. A comparison of the
predictions of this model with existing data and the expectations for future
colliders is presented.

\end{abstract}

%\vskip1.75in

\renewcommand{\thefootnote}{\arabic{footnote}} \end{titlepage}

%%%%%%%%%%%%%%%%%%%%%%%%%%%%%%%---- text

One of the hallmarks of the Standard Model (SM) is the universality of the
strong and electroweak interactions. Although well established experimentally
for the first two generations, the possibility of a small universality
violation (UV) for the third generation remains open{\cite{rolandi,marciano}}.
In fact, there is potentially some evidence which may be interpreted as a
signal for UV: the $\tau$-lifetime problem. Although the experimental
situation has evolved significantly over the past year{\cite {davier}}, the
possibility that $\tau$ decays may violate universality is still viable. If
such a UV were indeed confirmed it would be an unmistakable signal for new
physics beyond the SM.

The origin of any UV by the third generation could arise from a number of
sources. One intriguing possibility, proposed more than a decade ago, is the
model of Li and Ma{\cite {ma}}, in which each generation couples to its own
left-handed weak isospin group, $SU(2)_{\sigma}$, ($\sigma=1-3$) with its own
gauge coupling, $g_{\sigma}$. The $U(1)$ hypercharge group remains universal,
with a coupling $g_0$, and a global symmetry enforces the universality between
the first two generations. Clearly, in such a model, the {\it {size}} of the
UV in $\tau$ decays is not a prediction but rather an input parameter
with which we
can further analyze the implications of the model. For our purposes, we will
depart from the nomenclature of Li and Ma and define
\begin{equation}
\tau^{exp}/\tau\sub{SM} = (1+\epsilon)^2 \,.
\end{equation}
Experimentally, $\epsilon$ cannot be very large and two recent global analyses
give $\epsilon =(7.6\pm 6.6)\times 10^{-3}${\cite {davier}} and $\epsilon
=(13.2\pm 6.1)\times 10^{-3}${\cite {rolandi}}.

The purpose of this paper is to further explore the ramifications of this model
using these recently obtained values of $\epsilon$ as input. In particular,
we are interested in its implications for physics at LEP as well as other
$e^+e^-$ and hadron colliders. One possible approach is to treat $\epsilon$ as
a small parameter and expand all expressions for observables only to leading
order in $\epsilon$. The danger of this approach is that we can never be sure
that the coefficient of the next term in this expansion does not conspire to
invalidate our conclusions. To this end, we will examine the predictions of
the model numerically presenting lowest order expansions where appropriate. It
is important to stress that our analysis of the model relies solely on its
predictions of observable quantities. For the first two fermion generations,
and for the gauge boson sector, the deviations from the predictions of the SM
can be displayed by using the parameters introduced by Peskin and Takeuchi
{\cite {peskin}}:
$\Delta S$, $\Delta T$, and $\Delta U$, and which are already constrained by
existing data once SM radiative corrections are accounted for. The UV by the
third generation may be observed at LEP, \eg , by comparing the $Z \to \tau^+
\tau^-$ or $b \bar {b}$ widths with those expected on the basis of
universality. As we will see, for a given value of $\epsilon$, this data
already constrains the other parameters of the model. Let us first begin by
reviewing the basics aspects of the model that we need in our analysis.

At first sight this model would appear to suffer from a proliferation of
parameters but as we will now see there are actually only two new ones in
the limit that $e-\mu$ universality is preserved. The first, $\epsilon$, we
have already met above while a second can be defined in terms of the
low-energy effective neutral current Hamiltonian
\begin{equation}
{\cal H}\sub{NC} = {4G_F\over \sqrt 2} \left[ (J_{3L}-\xw J_{em})^2
+C(J_{em})^2 \right] \,.
\end{equation}
In this model, the parameter $C$ is highly constrained and we can in fact
define a new quantity, $p$, which is forced to lie on the unit interval:
\begin{equation}
0\leq p \equiv {1\over \xw} \sqrt{C\over\epsilon} \leq 1 \,.
\end{equation}
One sees from this Hamiltonian that $x_w$ is the effective mixing angle probed
by low energy $\nu$ data. At LEP or higher collider energies, we must
decompose this `composite' interaction into its `components' which arises from
the exchange of the individual $Z$ and $Z'$ gauge bosons:
\begin{eqnarray}
{\cal L}_{Z,Z'} &=& {-e\over\sqrt{x(1-x)}}(J_{3L}-xJ_{em})Z \\
& & \mbox{} + {e\over\sqrt{x}}\left[ {p(1+\epsilon)\over 1-p} \right]^{1/2}
\left( J_{3L}-{1+\epsilon p\over p(1+\epsilon)}J_{3L}\delta_{3\sigma}\right) Z'
\,, \nonumber
\end{eqnarray}
with $\sigma$ being a generation label and $e$ being the running electric
charge at, \eg , the $Z$ mass scale.
Here, an apparently new parameter has appeared, $x$, but we find that it is
directly related to $x_w$, $p$, and $\epsilon$:
\begin{equation}
x\equiv\xw(1+\epsilon p) \,.
\end{equation}
We can, of course, freely choose either $x$ {\it {or}} $x_w$ as the independent
parameter and write all of the various combinations of gauge couplings in
terms of this choice together with $p$ and $\epsilon$:
\begin{equation}
\begin{array}{ll}
e^2g_0^{-2} = x \,, \; & e^2g_3^{-2} = xp(1+\epsilon)/(1+\epsilon p) \,,
 \\
e^2g_{123}^{-2} = 1-x \,, \; & e^2g_{12}^{-2} = x(1-p)/(1+\epsilon p) \,.
\end{array}
\end{equation}
where we follow Li and Ma and employ the short-hand notation: $g_{123}^{-2}=
g_1^{-2}+g_2^{-2}+g_3^{-2}$ and similarly $g_{12}^{-2}=g_1^{-2}+g_2^{-2}$.
We find the parameter $x$ to be the more convenient choice for our
analysis. $Z-Z'$ mixing arises naturally from the $Z-Z'$ mass matrix
\begin{equation}
{\cal M}^2_{ZZ'} = {1\over 2} \left( \begin{array}{cc}
a_zv_3^2 \; & b_zv_3^2 \\
b_zv_3^2 \; & (c_z+d_z/\epsilon)v_3^2
\end{array} \right) \,,
\end{equation}
with $v_3$ being a vacuum expectation value and where the coefficients are
given by
\begin{equation}
\begin{array}{ll}
a_z=g_0^2+g_{123}^2 \,, \; & c_z=g_3^2-g_{123}^2 \,, \\
b_z=e^{-1}g_{12}^{-1}g_0g_3g_{123}^2 \,, \; & d_z=g_{12}^2+g_3^2 \,.
\end{array}
\end{equation}
The mass eigenstates are defined via the rotation
\begin{equation}
\left( \begin{array}{c}
Z \\ Z'
\end{array} \right) =
\left( \begin{array}{cc}
\cos\phi \; & \sin\phi \\
-\sin\phi \; & \cos\phi
\end{array} \right)
\left( \begin{array}{c}
Z_1 \\ Z_2
\end{array} \right) \,.
\end{equation}
The angle, $\phi$, is found through the simple relation
\begin{equation}
\tan 2\phi = {-2b_z\over c_z-a_z+d_z/\epsilon} \,,
\end{equation}
while $v_3$ is directly related to the Fermi constant
\begin{equation}
{G_F\over\sqrt 2} = {1+\epsilon\over 4v_3^2} \,.
\end{equation}
We note, for our analysis below, that the corresponding $W-W'$ mass matrix is
given by
\begin{equation}
{\cal M}^2_{WW'}={1\over 2} \left( \begin{array}{cc}
a_wv_3^2 \; & b_wv_3^2 \\
b_wv_3^2 \; & (c_w+d_w/\epsilon)v_3^2
\end{array} \right) \,,
\end{equation}
with
\begin{equation}
\begin{array}{ll}
a_w=g_{123}^2 \,, \; & c_w=c_z \,, \\
b_w=-g_{123}(g_3^2-g_{123}^2)^{1/2} \,, \; & d_w=d_z \,.
\end{array}
\end{equation}
It is important to note the relations $c_w=c_z$ and $d_w=d_z$ result from a
residual $SU(2)$ symmetry in the model. Because of these relationships and the
rather small size of $\epsilon$ the $W'$ and $Z'$ in this model are highly
degenerate.
For fermions in the first two generations, the couplings of the $Z_1$, the
state probed at LEP and SLC, can be written as
\begin{equation}
{\cal L}={-e\over\sqrt{x(1-x)}}N(J_{3L}-x\sub{eff}J_{em})Z_1 \,,
\end{equation}
with $x\sub{eff}$ being an `effective' mixing angle parameter and $N$ a
normalization correction both of which can be determined experimentally and
are given by
\begin{eqnarray}
x\sub{eff} &=& {x\cos\phi\over N} \,, \\
N &=& \cos\phi - \sqrt{1-x} \sin\phi\left[ {p(1+\epsilon)\over 1-p}
\right]^{1/2} \,. \nonumber
\end{eqnarray}
with $e=e(M_{Z_1})$ being understood.
The $Z_1$ mass can be obtained directly from the $Z-Z'$ mass matrix and can
written in the form
\begin{equation}
M^2_{Z_1} = {\pi\alpha(M_{Z_1})\over\sqrt{2}G_Fx(1-x)} [f(\epsilon)
-\epsilon^2g(\epsilon,x)] \,,
\end{equation}
where $f, g$ can be calculated to any desired order in $\epsilon$.
Unfortunately, unlike all other parts of our analysis, we cannot analytically
determine $x$ to all orders in $\epsilon$ from this equation. However, a
series solution is possible and can be generated with relative ease to
any desired order. Since
$\epsilon$ is relatively small, we find that this series expansion rapidly
converges numerically
especially since we find that it can be partially resummed. For example,
to order
$\epsilon^2$, and including this resummation, we can invert the $Z_1$ mass
relation above for $x$ and obtain
\begin{equation}
x=x\sub{r} \left[ 1-{2\epsilon^2g_0Ax\sub{r}\over (1-x\sub{r})(1-2x\sub{r})}
\right] \,,
\end{equation}
where to this order we have defined
\begin{eqnarray}
f &=& {1+\epsilon\over 1+\epsilon\left({1-p\over 1+\epsilon p}\right)^2} \,,
\nonumber \\
g_0 &=& {p(1-p)^3(1+\epsilon)^2\over (1+\epsilon p)^4} \,, \\
x\sub{r} &=& {1\over 2}\left[ 1-\sqrt{1-4fA} \right] \,, \nonumber
\end{eqnarray}
with $A=\pi \alpha(M_{Z_1})/({\sqrt {2}}G_F M_{Z_1}^2)$.
In our analysis presented below, we will analytically include all terms
of order $\epsilon^3$, together with partial resummation, with all the higher
order terms {\it {explicitly}} verified as being numerically insignificant as
can be shown from a complete iterative numerical solution.

In order to proceed with the calculation of the contributions to $\Delta T$,
$\Delta S$, and $\Delta U$ in this model, we follow the analysis of Peskin
and Takeuchi(PT). The first thing we do is define the usual auxiliary
quantity, $x_0$, which is given solely in terms of the observed $Z_1$ mass,
$G_F$, and $\alpha(M_{Z_1})$:
\begin{equation}
x_0(1-x_0)\equiv A
\end{equation}
Note that this is a {\it {model independent}} quantity and is used simply as
an input from experiment into our analysis. It does not, \eg , depend on the
shift in the $Z$ mass due to $Z-Z'$ mixing as claimed by an earlier analysis.
(The fact that it is independent of new physics is reason it was introduced by
PT.) Using the latest data from LEP as input{\cite {rolandi}, we obtain
$x_0=0.23136 \pm 0.00022$.
In terms of $x_0$, we can write $\Delta S$, $\Delta T$, and $\Delta U$ as
linear functions of three {\it {observables}} as given by PT:
\begin{eqnarray}
{M^2_{W_1}\over M^2_{Z_1}} - (1-x_0) &=& {\alpha(1-x_0)\over 1-2x_0}
\left[ -{1\over 2}\Delta S+(1-x_0)\Delta T+{(1-2x_0)\over 4x_0}\Delta U
\right] \,, \nonumber \\
x\sub{eff}-x_0 &=& {\alpha\over 1-2x_0}\left[ {1\over 4}\Delta S -x_0
(1-x_0)\Delta T \right] \,, \\
Z_*-1 &=& {\alpha\over 4x_0(1-x_0)}\Delta S \,, \nonumber
\end{eqnarray}
with $\alpha=\alpha(M_{Z_1})$ and where $Z_*$ is defined via the leptonic
decay width of the $Z_1$ as given by PT:
\begin{equation}
\Gamma_\ell=Z_*{\alpha(M_{Z_1})M_{Z_1}\over 48x\sub{eff}(1-x\sub{eff})}
\left[ 1+ (1-4x\sub{eff})^2\right] \left( 1+{3\alpha(M_{Z_1})\over 4\pi}
\right)\,.
\end{equation}
The three observable quantities on the left-hand side of Eq.(20) can
be calculated explicitly in terms of the model parameters, $p$ and $\epsilon$,
and $x_0$. While the $W_1$ to $Z_1$ mass ratio can be obtained directly from
the mass matrices above, $x_{eff}$ is given by Eq.(15) while $Z_*$, defined
via Eq.(21), can be written simply as
\begin{equation}
Z_*= {N^2x\sub{eff}(1-x\sub{eff})\over x(1-x)} \,.
\end{equation}
The PT parameters can thus be directly obtained and are shown as a function of
$p$ in Fig.~1 for the two central values of $\epsilon$ obtained by the
global analyses. Approximate, lowest order in $\epsilon$ expressions are
given by
\begin{eqnarray}
\Delta S &=& 4\epsilon p(1-p)x_0/\alpha \,, \nonumber \\
\Delta T &=& -\epsilon p^2/\alpha \,, \\
\Delta U &=& -4x_0\Delta T \,, \nonumber
\end{eqnarray}
and provide a reasonably adequate description of the results shown in the
figure. The important result to notice is that while $\Delta T$ is
{\it {negative}}, both $\Delta S$ and $\Delta U$ are positive. These results
differ somewhat from those previously obtained{\cite {ma2}} in the
literature as we rely {\it {solely}} on observables to define $\Delta S$,
$\Delta T$, and $\Delta U$. We remind the reader that the values we obtain
are due only to the new physics contained in this model and are over and above
those contributions arising from shifts in the SM radiative corrections
reference point, \ie , other choices of the top-quark and Higgs boson masses
($m_t$ and $m_H$, respectively). Of course, for a specific choice of these
quantities we can ask for the {\it {shift}} in various observables due to
the new physics contained in this model. Fig.~2 shows the shifts in $x_{eff}$,
$x_w$, the $W_1$ to $Z_1$ mass ratio
\begin{equation}
\delta\rho_0\equiv {M^2_{Z_1}\over (1-x_0)M^2_{W_1}} -1 \,,
\end{equation}
and the shift in the overall normalization of $Z_1$ partial width to lepton
pairs in the `$G_F$' scheme, $\rho_Z$, defined via
\begin{equation}
\Gamma_\ell\equiv\rho_Z {G_FM^3_{Z_1}\over 24\sqrt{2}\pi}
\left[ 1+(1-4x\sub{eff})^2\right] \left( 1+ {3\alpha(M_{Z_1})\over 4\pi}
\right) \,,
\end{equation}
as functions of $p$ for the same choice of $\epsilon$ values as in Fig.~1.
Recent analyses which attempt to determine the allowed ranges for the PT
parameters are generally found to favor negative values for both $\Delta S$
and $\Delta T$ while $\Delta U$ is hardly constrained by existing data
{\cite {para}}. We note that for small values of $p$, all of the PT parameters
are quite small while for large $p$, \eg, $p \geq 0.6$, $\Delta S$ is small
and positive while $\Delta T$ can be large and negative. We will see that this
large $p$ region is particularly interesting when the UV experienced by the
third generation are considered and to this we now turn. By combining the
constraints from limits on UV together with those from an analysis of the PT
parameters we will find that the the model of Li and Ma is now highly
constrained and that $\epsilon$ is forced to be small.

{}From the results in Figs.~1 and 2 it would appear that in the $p \to 0$ limit
one would recover the predictions of the SM. This is in fact true {\it {only
for the first two generations}}. Eq.(4) tells us that the additional couplings
of the $Z'$ for the third generation {\it {grow}} with decreasing $p$ and thus
we expect to see the largest effect from UV in the $p \to 0$ limit. This is
shown explicitly in Fig.~3 where we display the fractional change in both the
$Z \to b \bar b$ and $Z \to \tau ^+\tau ^-$ partial widths from the
expectations of universality as functions of $p$. Analytically, these
deviations from universality for both these decay modes can be expressed as
\begin{equation}
{N_3^2[1+(1-4x_3)^2]\over N^2[1+(1-4x\sub{eff})^2]} -1
\end{equation}
for the case of taus(with $x_{eff}\to x_{eff}/3$ and $x_3 \to x_3/3$ ~in
the $b$ case),
where we define, in a manner similar to what we did above for the first two
generations, the third generation effective coupling parameters, $N_3$ and
$x_3$, that are given by
\begin{eqnarray}
x_3 &=& {x\cos\phi\over N_3} \,, \nonumber \\
N_3 &=& \cos\phi + \sin\phi\sqrt{1-x} \left[ {1-p\over p(1+\epsilon)}
\right]^{1/2} \,.
\end{eqnarray}
While sufficiently precise
data is not yet available on the $Z \to b \bar b$ partial width, such data
does exist
in the $\tau$ case{\cite {rolandi,davier}}. Correcting for the finite mass of
the $\tau$, we find that the deviation from universality can be at most $-1.26
\%$ at the $95\%$~CL. This would rule out values of $p$ below 0.22(0.55) for
the smaller(larger) choice of $\epsilon$ used in this analysis. Of course this
implies larger deviations from the SM for the first two generations and, as
discussed above, is just the region where $\Delta T$ can be large and negative
while $\Delta S$ remains small and positive. Fig.~4 shows the region excluded
in the $p-\epsilon$ plane using the constraints from LEP on violations of
$\tau$ universality. To go further we must make some assumptions about the
usual SM radiative corrections.

To be definitive, let us assume that the SM Higgs mass is 250 GeV with a top
quark mass of 120(150, 180) GeV so that SM radiative corrections can be
performed
completely. If we then calculate $x_3$ in the SM (which is just the usual
$sin^2 \theta_{eff}$) and compare with the LEP data
{\cite {rolandi}} for the $\tau$ polarization and forward-backward asymmetry
as well as the corresponding $b$ quark asymmetry, we find that another
sizeable region of the $p-\epsilon$ plane is now excluded at the $95\%$ CL
as shown in Fig.~4. Also shown in the figure are the corresponding regions
which are excluded by a recent $\Delta T - \Delta S$ analysis {\cite {para}}
with $m_t$ now fixed at 150 GeV. As can be seen, a combination of all these
constraints results in the requirement that $\epsilon$ must be reasonably
small for all values of $p$ and thus favoring the smaller value obtained in
the global fits of $\tau$ data.

In addition to improved $Z_1$ and $\tau$ data, this model can be probed at
hadron colliders by searches for the $Z'$ and $W'$. Given a set of fixed
values for $p$ and $\epsilon$, the $Z-Z'$ and $W-W'$ mass matrices can be
used to determine the masses of these new gauge bosons. To a excellent
approximation, the $W'$ and $Z'$ are degenerate with masses given by
\begin{equation}
M_{Z',W'}^2/M_{Z}^2 = {(1-x)\over {\epsilon p(1-p)}}
\end{equation}
and is shown in Fig.~5 without the use of any approximations. Taking the more
conservative determination of $\epsilon$ from the global fits to $\tau$ data
we see that the $Z'$ and $W'$ must have masses in excess of 1 TeV at the
$95\%$ CL. We can, of course, ask what limits can be set on the $Z'$ and $W'$
from Tevatron data which is presently available, or will be available before
the turn-on of the SSC/LHC; this is shown in Fig.~6. As we can see from this
figure, the Tevatron will be barely able to touch the region of interest even
if an integrated luminosity as large as $1~fb^{-1}$ is accumulated. The $Z'$
and $W'$ physics in this model must be left to the SSC/LHC or TeV $e^+e^-$
linear colliders to probe. Fig.~7, for example, shows the anticipated signal
rate for the $Z'$ in this model at the SSC after cuts and efficiencies are
accounted for assuming an integrated luminosity of 10 $fb^{-1}$, corresponding
to 1 `SSC-year.' Certainly a $Z'$ in the several TeV mass range will prove to
be easily observable. In the near future, however, it is thus most likely
that stricter constraints upon the parameter space of this model will only
result from further refinements in $\tau$ and $Z_1$ data, {\it {provided}}
universality violations are not in fact observed.

In this paper we have re-examined the predictions of the
universality violating model of Li and Ma in light of recent high precision
data from LEP on the properties of the $Z$ as well as new data on the
properties of the $\tau$. We find that a combination of this data, taken
together with a knowledge of the SM radiative corrections, places very strong
constraints on the two parameters of the model: $p$ and $\epsilon$. For a top
quark mass of 150 GeV and a Higgs scalar mass of 250 GeV, we find for example
that values of $\epsilon >0.009$ are excluded for all values of $p$, while for
$p$ less than 0.3, even small values of $\epsilon$ are excluded. Existing
constraints on the model parameters were shown to forbid the direct discovery
of the new $Z'$ and $W'$ gauge bosons at the Tevatron but they may be
copiously produced at, \eg , the SSC if their masses do not exceed a few TeV.

Searches for the possibility of universality violation by the particles in
the third generation must continue.

\vskip.25in
\centerline{ACKNOWLEDGEMENTS}

The author would like to thank E. Ma and J.L. Hewett for extensive discussions
related to this work.
This research was supported in part by the U.S.~Department of Energy
under contract W-31-109-ENG-38.

\newpage

%
%%%%%%%%%%%%%%%%%%--- References
%%%%%%%%%%%%%%%%%%%%%%%%%%%%%%%%%%%%%%%%%%%%%%%%%%%%%%%
\def\MPL #1 #2 #3 {Mod.~Phys.~Lett.~{\bf#1},\ #2 (#3)}
\def\NPB #1 #2 #3 {Nucl.~Phys.~{\bf#1},\ #2 (#3)}
\def\PLB #1 #2 #3 {Phys.~Lett.~{\bf#1},\ #2 (#3)}
\def\PR #1 #2 #3 {Phys.~Rep.~{\bf#1},\ #2 (#3)}
\def\PRD #1 #2 #3 {Phys.~Rev.~{\bf#1},\ #2 (#3)}
\def\PRL #1 #2 #3 {Phys.~Rev.~Lett.~{\bf#1},\ #2 (#3)}
\def\RMP #1 #2 #3 {Rev.~Mod.~Phys.~{\bf#1},\ #2 (#3)}
\def\ZP #1 #2 #3 {Z.~Phys.~{\bf#1},\ #2 (#3)}
\def\IJMP #1 #2 #3 {Int.~J.~Mod.~Phys.~{\bf#1},\ #2 (#3)}

\newpage

%%%%%%%%%%%%%%%%%%%%%%%--- figures
%
{\bf Figure Captions}
\begin{itemize}

\item[Figure 1.]{(a)$\Delta T$, (b)$\Delta S$, and (c)$\Delta U$ as functions
of the parameter $p$ with the smaller(larger) value of $\epsilon$ discussed in
the text corresponding to the dotted(dashed) curve.}
\item[Figure 2.]{Shifts in the values of (a)$x_{eff}$, (b)$\delta \rho_0$,
(c)$x_w$,
and (d)$\delta \rho_Z$ as functions of $p$ as described in the text for the
same two
choices of $\epsilon$ as shown in Fig.~1.}
\item[Figure 3.]{Fractional change in the (a)$Z \to b \bar b$ and (b)
$Z \to \tau^+ \tau^-$ partial widths as functions of $p$ for the two choices
of $\epsilon$
shown in Fig.~1 relative to the predictions of universality.}
\item[Figure 4.]{Allowed region in the $p-\epsilon$ plane from LEP data on
the $Z\to
\tau^+ \tau^-$ partial width(solid) and from a radiative corrections analysis
determination of the bound on $x_3$ assuming a SM Higgs mass of 250 GeV and a
top quark mass of 120(leftmost), 150(center), or 180(rightmost) GeV
represented by the sequence of dashed curves. The allowed region is to the
left of each curve. The corresponding bounds from a $\Delta S$(dotted) and
$\Delta T$(dash-dots) analysis is also shown for the $m_t$=150 GeV case.}
\item[Figure 5.]{Lower bound on the $Z'$ or $W'$ masses in this model as
functions of the parameter $p$ for the same choices of $\epsilon$ as in
Fig.~1.}
\item[Figure 6.]{Tevatron search limits for the $Z'$ or $W'$ as functions of
$p$: present limits(dotted), and future limits from a total integrated
luminosity of 25(100, 400, 1000) $pb^{-1}$ shown as a dashed(dash dotted,
solid, square-dotted) curve.}
\item[Figure 7.]{Number of $Z'$-induced dilepton events which would be
observed at the SSC, as a function of $p$, assuming an integrated luminosity
of 10 $fb^{-1}$, a lepton
identification efficiency of 0.85, and demanding both leptons lie in the
pseudorapidity interval $-2.5 \leq \eta \leq 2.5$. The $Z'$ mass is assumed
to be 1(2, 3, 4, 5) TeV in the case of the dotted(dashed, dash-dotted, solid,
square-dotted) curve.}
\end{itemize}


\begin{thebibliography}{99}
\bibitem{rolandi}
L.\ Rolandi,
plenary talk given at the {\it {XXVI ICHEP 1992}}, Dallas, Texas, August 1992.
\bibitem{marciano}
W.\ Marciano,
plenary talk given at the {\it {Division of Particles and Fields Meeting}},
Fermilab, November 1992.
\bibitem{davier}
M.\ Davier,
summary talk of the {\it {Second Workshop on Tau Lepton Physics}}, The Ohio
State University, Columbus, Ohio, September 1992.
\bibitem{ma}
X.-Y.\ Li and E.\ Ma, \PRL 47 1788 1991 ;
E.\ Ma and D.\ Ng, \PRD D38 304 1988 ;
X.-Y.\ Li and E.\ Ma, \PRD D46 1905 1992 ;~and University of
California-Riverside report UCRHEP-T97 (1992).
\bibitem{peskin}
M.E.\ Peskin and T.\ Takeuchi, \PRD D46 381 1992 ;~ \PRL 65 964 1990 .
\bibitem{ma2}
See the last two papers listed in Ref.~4.
\bibitem{para}
For a recent updated analysis, see P.\ Roy, talk given at the {\it {X DAE High
Energy Physics Symposium}}, Tata Institute, Bombay, December, 1992;
see also, for example, G.\ Bhattacharya, S.\ Banerjee and P.\ Roy,
\PRD D45 R729 1990  ~and erratum \PRD D46 3215 1992 .

\end{thebibliography}
\end{document}